\documentstyle[preprint,revtex]{aps}
\newcommand{\be}{\begin{equation}}
\newcommand{\ee}{\end{equation}}
\newcommand{\bea}{\begin{eqnarray}}
\newcommand{\eea}{\end{eqnarray}}
\begin{document}

\begin{title}
Narrow band noise as a model of time-dependent accelerations: study of
the stability of a fluid surface in a microgravity environment
\end{title}

\author{Jaume Casademunt, Wenbin Zhang, Jorge Vi\~nals}
\begin{instit}
Supercomputer Computations Research Institute, B-186,
Florida State University, Tallahassee, Florida 32306-4052.
\end{instit}

\moreauthors{Robert F. Sekerka}
\begin{instit}
Departments of Physics and Mathematics,
Carnegie Mellon University, Pittsburgh, Pennsylvania 15213-3890.
\end{instit}

\begin{abstract}
We introduce a stochastic model to analyze in quantitative detail the
effect of the high frequency components of the residual accelerations
onboard spacecraft (often called g-jitter) on fluid motion. The residual
acceleration field is modeled as a narrow band noise
characterized by three independent parameters: its intensity $G^{2}$, a
dominant
frequency $\Omega$, and a characteristic spectral width $\tau^{-1}$.
The white noise limit corresponds to $\Omega \tau \rightarrow 0$, with
$G^{2} \tau$ finite, and the limit of a periodic g-jitter (or
deterministic limit) can be recovered for $\Omega \tau \rightarrow
\infty$, $G^{2}$ finite. The analysis of the response of a fluid surface
subjected to a fluctuating gravitational field
leads to the stochastic Mathieu equation driven by both additive and
multiplicative noise. We discuss the stability of the solutions of this
equation in the two limits of white noise and deterministic forcing, and
in the general case of narrow band noise. The results are then applied
to typical microgravity conditions.
\end{abstract}

\narrowtext

\section{Introduction}

A stochastic model is introduced to describe the
high-frequency components of the residual accelerations onboard
spacecraft (often called g-jitter). The model is incorporated into the
equations governing fluid motion, and the stability of a surface of
discontinuity between two fluids of different density is analyzed.
The linear stability of this surface is governed by the
stochastic Mathieu equation, driven by both multiplicative and additive
noise. We study the range of parameters in which the solutions of the
stochastic Mathieu equation are stable, and apply the results obtained
to a water-air surface in typical microgravity conditions. We find that
the component of g-jitter normal to the surface at rest couples
nonlinearly to the surface displacement and can lead to parametric
instability. On the other hand, the two parallel components appear
additively and we show that they do not modify the stability boundaries
associated with the normal component.

A certain amount of attention has been paid recently to modeling
g-jitter as a periodic function of time \cite{re:wa87,re:al90,re:ne91},
but little attention has been
paid to the more realistic case in which the effective acceleration
spectrum contains a band of frequency components and is random in
nature. Early work in this direction by Antar \cite{re:an77},
considered the stability
of the Rayleigh-B\'enard configuration under a random gravitational
field with a uniform frequency spectrum (white noise). Later Fichtl and
Holland used a stochastic description to study the distribution of
impulses that would exceed a prescribed threshold \cite{re:fi78}.

We introduce a general model for g-jitter, also based on a
stochastic description \cite{re:vi90}, with a spectrum that is similar
to residual
acceleration spectra measured in space missions. The model, which we
call narrow band noise, is characterized by three parameters: $G^{2}$,
the mean intensity of the fluctuations, $\Omega$, their characteristic
frequency, and $\tau^{-1}$, the characteristic width of the spectrum
(peaked at $\Omega$) \cite{re:zh92}. In the limit
$\Omega \tau \rightarrow 0$,
with $D = G^{2} \tau$ finite, narrow band noise reduces to white noise
of intensity $D$, whereas for $\Omega \tau \rightarrow \infty$ and
$G^{2}$ finite, monochromatic noise of intensity $G^{2}$ is recovered.
Monochromatic noise is equivalent to a deterministic periodic forcing of
circular frequency $\Omega$ and amplitude $G/\sqrt{2}$.
{}From published acceleration
spectra measured during space missions \cite{re:al90,re:ne91}, one can
infer that characteristic frequencies of g-jitter are
in the range $1-20~Hz$, hence $\Omega \sim 2 \pi (1-20) s^{-1}$.
Characteristic widths of the spectral density are $5-10~Hz$, or $\tau
\sim 0.1 - 0.2 s$. Therefore $\Omega \tau \sim 1~-~25$, which is an
intermediate case between the white noise limit and the deterministic
limit of periodic forcing.

The motivation for our work is that high-frequency components of the
residual acceleration field will affect fluid flow primarily in regions
where large gradients of density exist. The case of a surface of
discontinuity studied in this paper is a prototypical example, and is
used to explore in detail the coupling between a time dependent
acceleration and fluid motion due to density gradients. Since our
analysis leads to the stochastic Mathieu equation, our results are
applicable to a variety of situations, including parametrically
excited surface waves \cite{re:mi90},
and convective instabilities in systems under a
time-dependent gravitational field \cite{re:wh91}.

Our results show that instability in the stochastic case is mainly due to
subharmonic parametric resonance. There is, however, a complex
interplay between the linear dispersion relation of the surface and the
power spectrum of the external acceleration field which results in two
main differences with respect to
the classical deterministic case of periodic forcing.
First, stochastic forcing excites a range of frequencies. Parametric
resonance can occur for modes which would not be resonant
with the exciting frequency in a deterministic model. Second, stability is
typically enhanced when moving away from the deterministic limit into the
narrow band noise region.

\section{Equation governing interface displacements}

Consider two incompressible, immiscible fluids in a
finite container, initially separated by a planar surface at $z=0$.
We assume that fluid 1 of density $\rho_{1}$ and shear viscosity $\mu_{1}$
occupies the region $-d < z < 0$, and fluid 2 of density $\rho_{2}$ and
shear viscosity $\mu_{2}$ occupies the region $0< z <d$.  We summarize here
the conditions under which the fluid equations can be reduced to a closed
equation for the interface displacement alone when the system is subjected to
arbitrary, time dependent accelerations
$\vec{g}(t) = (g_{x}(t), g_{y}(t), g_{z}(t)-g_{0})$. $g_{0}$ is the
static component of the acceleration directed along the $z$-axis.
As discussed in Ref. \cite{re:zh92}, such an equation only exists
when the time scales associated with viscous dissipation
are much longer than typical time scales of
interface motion (i.e., in the underdamped limit). In this regime,
fluid flow can be assumed to be potential
except in thin boundary layers at the container walls and around
the fluid interface. We will neglect in what follows any
capillary effects at the contact between the interface and the container wall.

Following the formulation of Miles \cite{re:mi76} for the case of a single
inviscid fluid, we expand both the interface displacement $h(x,y,t)$
away from its reference position, and the velocity potential
$\phi_{i}(x,y,z,t)$, as
\bea
h(x,y,t) &=& \sum_{n}h_{n}(t)\psi_{n}(x,y),  \\
\label{eq:phi}
\phi_{i}(x,y,z,t) &=& \sum_{n}\phi_{in}(t)\psi_{n}(x,y)Z_{in}(z),
\eea
where $\psi_{n}(x,y)$ are the eigenfunctions of
$(\nabla^{2}+k_{n}^{2})\psi_{n}(x,y) = 0$ with boundary condition
$\hat{n}\cdot\nabla\psi_{n}=0$ at the container's wall.
The equation for $\psi_{n}$ follows from Laplace's equation for the
velocity potentials, $\phi_{i}$, when their $z$ dependences are solved for
each eigenmode as $Z_{1n}=\cosh k_{n}(z+d)/\cosh k_{n}d$ and
$Z_{2n}=\cosh k_{n}(z-d)/\cosh k_{n}d$, where $d$ is the depth of the
fluid layer, assumed to be equal in both fluids for simplicity.
The eigenfunctions are orthogonal
and normalized, $\int\!\!\int\psi_{m}\psi_{n}dxdy = S\delta_{mn}$,
with $S$ the cross-sectional area of the container.
If cubic and higher orders terms in $h_{n}$ and $\dot{h}_{n}$ are neglected,
the Lagrangian function for the generalized coordinates $h_{n}(t)$
reads,
\be
L = \sum_{n}(\rho_{1}+\rho_{2})S\left(\frac{\dot{h}_{n}^{2}}{2k_{n}'}
- \frac{1}{2}A(g_{0}-g_{z}(t))h_{n}^{2}
+ A Q_{n}(t)h_{n}
- \frac{\Gamma k_{n}^{2}h_{n}^{2}}{2(\rho_{1}+\rho_{2})} \right),
\ee
where $k_{n}' = k_{n}\mbox{tanh}(k_{n}d)$,
$A = (\rho_{1}-\rho_{2})/(\rho_{1}+\rho_{2})$, $\Gamma$ is the interfacial
tension, and
\be
\label{eq:Qn}
Q_{n}(t) = \frac{1}{S}\int\!\!\int(g_{x}(t)x+g_{y}(t)y)\psi_{n}dxdy.
\ee

Viscous dissipation may now be approximately
incorporated by calculating the rate of energy dissipation ${\cal D}$ of the
fluid system.  By introducing the dissipation function ${\cal F} = -
{\cal D}/2$, we have the following equation governing the motion of the
viscous fluid,
\be
\label{eq:Lagr}
\frac{d}{dt}\left(\frac{\partial L}{\partial \dot{h}_{n}}\right) -
\frac{\partial L}{\partial h_{n}} +
\frac{\partial{\cal F}}{\partial \dot{h}_{n}} = 0.
\ee
If the dissipation function ${\cal F}$
can be written, in some approximation, as
\be
\label{eq:Fform}
{\cal
F}=\frac{1}{2}
S(\rho_{1}+\rho_{2})\sum_{n}\frac{\gamma_n}{k_{n}'}\dot{h}_{n}^{2},
\ee
then Eq.~(\ref{eq:Lagr}) reads,
\be
\label{eq:hn}
\frac{d^{2}h_{n}}{dt^{2}} + \gamma_{n}\frac{dh_{n}}{dt}
+ (\omega_{n}^{2} - Ak_{n}'g_{z}(t))h_{n} =
Ak_{n}'Q_{n}(t),
\ee
with,
\be
\label{eq:disrel}
\omega_{n}^{2} = \frac{(\rho_{1}-\rho_{2})g_{0}k_{n}'+\Gamma k_{n}^{2}k_{n}'}
{\rho_{1}+\rho_{2}}.
\ee
The total viscous dissipation can be split into three different contributions,
namely, dissipation in the bulk, at the interface boundary layers and
at the container's walls.
As shown in the Appendix, the dissipation due
to potential flow in the bulk, in the linear approximation and for
a rectangular cross section of the container, can be written in the form
of Eq. (\ref{eq:Fform}), with
$\gamma_n = 4 \frac{\mu_1 + \mu_2}{\rho_1 + \rho_2}k_n^2$.
Therefore, if dissipation at the boundary layers could be neglected
compared to bulk dissipation, a closed equation of the form Eq. (\ref{eq:hn})
can be obtained for the generalized coordinates describing the
displacement modes of the interface.
For the two fluid case, however, the ratio of
interface dissipation to bulk dissipation is of order
$\sim \sqrt{\omega_n/(\nu_1+\nu_2)k_{n}^{'2}}$,  which is a large
quantity in the underdamped limit (see the Appendix).
Hence a closed equation like Eq. (\ref{eq:hn}) is not satisfied.
As shown in Ref. \cite{re:zh92}
for a single fluid (i.e., $\rho_2=0$, $\mu_2=0$), bulk dissipation
does dominate over interface dissipation and Eq. (\ref{eq:hn}) holds.
A closed equation for the generalized coordinates $h_{n}(t)$ can also be
obtained for a system of finite size if the dissipation at the
walls of the container is negligible compared to bulk dissipation.
The ratio of bulk to wall dissipation can be estimated as
$\sim Lk_{n}^{2}\sqrt{2\nu/\omega_{n}}$ (see the Appendix).
Therefore, for a given driving frequency range, and the concomitant
resonant modes  $k_n$, Eq. (\ref{eq:hn}) with $\gamma_n=4\nu k_n^2$
is recovered provided the system size
is large enough. For a water-air interface,
and a typical g-jitter frequency of $\sim 10 Hz$, bulk dissipation
dominates for $L > 5 cm$ (typically, the condition of large bulk to wall
dissipation ratio is satisfied if $L$ is larger than the wavelength of
the surface modes).

Equation (\ref{eq:hn}) is our starting point for the stochastic analysis
presented in the next section.
The salient feature of Eq.~(\ref{eq:hn}) is that the contribution of
$g_{z}(t)$ is nonlinear or
multiplicative ($g_{z}(t) h_{n}(t)$), whereas the contributions from $g_{x}(t)$
and $g_{y}(t)$ are additive. It can also be shown that, contrary
to the multiplicative forcing, the strength
of the additive forcing is size dependent and vanishes in the limit of an
infinite cross-sectional area.  For instance, in a rectangular container,
Eq.~(\ref{eq:Qn}) with the explicit form of the
corresponding eigenfunctions, can be written as,
\be
Q_{ij}(t)= \frac{\sqrt{2}}{k_{ij}^{2}}
\left(\frac{((-1)^{i}-1)g_{x}(t)}{L_{x}}\delta_{j0} +
\frac{((-1)^{j}-1)g_{y}(t)}{L_{y}}\delta_{i0}\right).
\ee
In the limit of an infinite
container, $L_{x}, L_{y} \rightarrow \infty$, $Q_{ij}(t) \rightarrow 0$
for any fixed wavenumber $k_{ij}$.

\section{Oscillator driven by multiplicative and additive noise}

We have seen that under certain conditions, each normal
mode of the interface displacement satisfies a closed equation which
is equivalent to that of
a linear oscillator subject to both  multiplicative and  additive
forcing. The intensity of the
additive forcing depends on system size, but not on the actual interface
displacement, and vanishes in the limit of large system size. The
intensity of the multiplicative
term is independent of system size, but depends on the actual displacement
of the interface, and vanishes in the reference state of a planar interface at
$z=0$.

In this section we study the statistical properties of a dynamical
variable $x(t)$ that satisfies the equation,
\begin{equation}
\label{eq:osc}
\frac{d^2 x}{d t^2} + \gamma\frac{dx}{dt} + \omega_0^2x + \xi(t)x = \zeta(t),
\end{equation}
where $\xi(t)$ and $\zeta(t)$ are two arbitrary stochastic forces.
We characterize the stability of the solutions in terms of the stability
of the
statistical moments of $x(t)$ as a function of the parameters of
the system ($\gamma$ and $\omega_{0}$) and of the noise. We restrict our
study here to the second order moments, $<x^2>$, $<x\dot{x}>$ and
$<\dot{x}^2>$,
that are directly proportional to the energy of the oscillations.

\subsection{Uncorrelated white noises}

If the two stochastic forces in Eq.(\ref{eq:osc}) are white and
uncorrelated, $ <\xi(t)\xi(t')>=2D\delta(t-t')$,
$<\zeta(t)\zeta(t')>=2\epsilon\delta(t-t')$, and
$ <\xi(t)\zeta(t')>=0$,
the evolution equation for the
second order moments is closed and linear. The so called energetic
instability occurs when one of the three eigenvalues
of the evolution matrix for the second order moments, say $\lambda_{1}$ has
a positive real part, implying an exponential growth of the second moments.
 This eigenvalue is known exactly \cite{re:li90}
and, in the underdamped limit, $\gamma^{2}/4 \omega_{0}^{2}
\ll 1$,  reads,
\be
\lambda_1 = -\gamma + \frac{D}{\omega_0^2},
\ee
and is independent of the intensity of the additive noise $\epsilon$.
The white noise limit corresponds to the idealized situation in which
the spectrum of the stochastic force is constant. For more realistic
noise spectra, the evolution equations for the
second order moments are not closed in general, and in order to define a
similar criterion of instability, different truncation schemes have to
be invoked.

\subsection{Deterministic periodic forcing}

The opposite limit to white noise corresponds to a periodic forcing
of the form
$\xi(t)=\xi_0\cos{\Omega t}$ and $ \zeta(t)=\zeta_0\cos{(\Omega t + \phi_0)}$.
This limit models a monochromatic g-jitter of frequency $\Omega$ with
highly correlated components.
In this limit, the stability properties of Eq. (\ref{eq:osc})
can be determined as follows. In the absence of the additive term ($\zeta_0=0$)
Eq. (\ref{eq:osc}) reduces to the classical (damped) Mathieu equation.
For $\gamma=0$, the solutions are always stable and
periodic for small enough $\xi_0$, except at the so-called parametric
resonances that
occur when $\omega_0=n\Omega/2$, with $n=1,2, \ldots$ The strongest
resonance occurs at $n=1$. For finite $\gamma$, however, there is a finite
threshold for instability,
\be
\label{eq:detthres}
\left( \frac{\xi_{0}}{2 \omega_{0}} \right)^{2} - \gamma^{2} \ge 0.
\ee
We argue that the regions of stability of the solutions of the damped
Mathieu equation do
not to depend on an
additive periodic forcing of the same period. This result follows from
general theorems of linear differential equations with
periodic coefficients. According to 2.9 in
Ref. \cite{re:ya75}, if the homogeneous part of Eq. (\ref{eq:osc}) does not
have
periodic
solutions, and the inhomogeneous term is itself periodic with the same
period as the coefficients of the homogeneous part, then the particular
solution of the full equation is itself a periodic function. Therefore the
stability of the solutions of the full equation
remains unchanged after the addition of the
inhomogeneous terms, since instability will only occur if the
homogeneous part already has a positive Floquet exponent.
In our case, the Floquet exponents of the homogeneous part
$\alpha_1, \alpha_2$ (i.e., of the damped Mathieu equation)
are related to those of the undamped
case, $\pm \sigma$, by $\alpha_{1,2}=-\gamma/2 \pm \sigma$,
(see 4.3 in \cite{re:ya75}). Therefore, the
solutions of the homogeneous equation
are not periodic except when $\gamma = 2 \sigma$. But this
is precisely the condition defining the stability boundary of the damped
Mathieu equation. Therefore, the stability boundaries of
Eq. (\ref{eq:osc}), for the deterministic case, are also insensitive to
additive forcing.

\subsection{Narrow band noise}

We now consider the generic case of a narrow band noise. We have shown that
the additive contribution to Eq. (\ref{eq:osc}) does not modify the
stability boundaries of the second order moments in the two limits of
white noise and deterministic forcing. It is natural to expect that
in the  intermediate case of narrow-band noise, which interpolates
between the two limits, the presence of an
additive noise, even if highly correlated to the multiplicative noise,
will not affect the stability boundaries either.
Therefore we restrict our stability study for narrow band noise to the
multiplicative contribution in Eq. (\ref{eq:osc}).
We define the spectrum of narrow band noise as,
\be
P_\xi(\omega) = \frac{1}{2\pi}<\xi^2>\tau \left(\frac{1}
{1+\tau^2(\Omega+\omega)^2} + \frac{1}
{1+\tau^2(\Omega-\omega)^2} \right).
\ee
This spectrum corresponds to a correlation function of the form,
\be
<\xi(t)\xi(t')>=<\xi^2> e^{-|t-t'|/\tau} \cos{\Omega(t-t')}
\ee
The limit $\Omega \tau \rightarrow 0$, $< \xi^{2} > \tau = D$ finite,
corresponds to white noise of intensity $D$. In the opposite limit,
monochromatic noise or, equivalently, a periodic forcing of amplitude
$\xi_{0}$ is recovered with $< \xi^{2} > = \xi_{0}^{2}/2$.

We have investigated this case by perturbation theory and
by numerical means \cite{re:zh92}. We have shown
that the underlying mechanism of
instability in the stochastic case can also be understood as a
parametric instability. For a wide range of parameters of the noise,
it is shown in \cite{re:zh92} that the eigenvalue describing
the stability of the second moments is well approximated by the expression,
\be
\label{eq:ansatz}
\lambda_1 \simeq -\gamma + \frac{1}{\omega_0^2} \pi P_\xi(2\omega_0).
\ee
The dependence of the instability threshold on the value of the power
spectrum at precisely $2\omega_0$  reflects the fact that the
underlying mechanism for instability can still be understood as
subharmonic parametric resonance.

\subsection{Effect of additive noise}

Additive noise in Eq.(\ref{eq:osc}) simply
modifies the spectrum of excitations of the interface in the cases
in which the interface is stable. For $\lambda_1>0$ the average energy
of the oscillations diverges exponentially in time,
until nonlinearities saturate the growth. The presence of additive
noise is irrelevant. On the other hand, for $\lambda_1<0$ additive
noise determines a finite steady value for the second
moments, which would vanish if only multiplicative noise were present.

In order to discuss the influence of the additive term on
the steady state fluctuations of the interface, let us first consider
Eq. (\ref{eq:osc}) without multiplicative noise ($\xi(t) = 0$).
Equation (\ref{eq:osc}) becomes linear, and the effect of a noise comprising
a band of frequency components can be described as a linear superposition of
the solution of the oscillator forced by a single frequency component.
The power spectrum $P(\omega)$ of the process $x(t)$ is defined
in the stationary regime as \cite{re:ga85},
\be
\label{eq:ps}
P(\omega)=\frac{1}{2\pi}\int_{-\infty}^{\infty} e^{-i\omega s}
<x(t)x(t+s)>ds .
\ee
Equation (\ref{eq:osc}) with $\xi(t) = 0$ reads, in Fourier space,
\be
\label{eq:oscFourier}
-\omega^2\hat{x}(\omega) + i\omega \gamma \hat{x}(\omega) + \omega_0^2
\hat{x}(\omega) = \hat{\zeta}(\omega).
\ee
By using Eq. (\ref{eq:oscFourier}) and the relations \cite{re:ga85}
$ <\hat{x}(\omega)\hat{x}^*(\omega')> = \delta(\omega-\omega') P(\omega)$,
and $<\hat{\zeta}(\omega)\hat{\zeta}^*(\omega')> = \delta(\omega-\omega')
P_\zeta(\omega)$,
we have,
\be
P(\omega) = \frac{P_\zeta(\omega)}{\left[\omega_0^2-\omega^2\right]^2 +
 \gamma^2 \omega^2},
\ee
where $P_\zeta(\omega)$ is the power spectrum of the additive noise.
Finally, by using Eq. (\ref{eq:ps}) the second moment $<x^2>$ can be written
as $<x^2> = \int_{-\infty}^{\infty} P(\omega) d\omega$.
In the deterministic limit, we have
$P_\zeta(\omega)=\frac{1}{2}\zeta_0 \delta(\omega-\Omega)$
and,
\be
<x^2>=\frac{1}{2} \frac{\zeta_0}{\left[\omega_0^2-\Omega^2\right]^2 +
 \gamma^2 \Omega^2},
\ee
where the average is now understood over a period in the stationary regime.
For white noise, we have $P_\zeta(\omega)=\frac{\epsilon}
{\pi}$ and,
\be
<x^2>=\frac{\epsilon}{2\gamma\omega_0^2}.
\ee
For narrow-band noise, the second moment can be expressed as,
\be
<x^2> = \frac{<\zeta^2>\tau}{2\pi} \int_{-\infty}^{\infty}
\frac{d \omega}{\left[\omega_0^2-\omega^2\right]^2 + \gamma^2 \omega^2}
 \left(\frac{1}{1+\tau^2(\Omega+\omega)^2} +
       \frac{1}{1+\tau^2(\Omega-\omega)^2} \right).
\ee
The excitation mechanism in all these cases
is an ordinary resonance phenomenon in contrast with
the parametric resonance of the multiplicative case. In both the
multiplicative and additive cases,
the variable $x$ responds in time with the
characteristic frequency of the forcing, $\Omega$, but in the purely additive
case the oscillations have a finite amplitude which depends on $\Omega$ and the
characteristic frequency of the
oscillator $\omega_0$. Unlike the multiplicative case,
the amplitude of oscillation saturates due to the dissipative forces
even within the linear regime, and instability in the sense of exponential
growth of the second moments cannot occur.

\subsection{Higher order moments}

In the general case in which both noise contributions are present but the
system is below threshold for parametric instability, the distribution
of fluctuations of the interface displacement cannot be determined exactly
for narrow band noise. In the limit in which both noises are white,
it has been shown in ref. \cite{re:li90} that the distribution $P(E)$ of
the energy of the oscillations
$E=\frac{1}{2}\left(\dot{x}^2+\omega_0^2x^2\right)$ (in the envelope
approximation, which assumes that the time scale for change of $E$ is much
larger than that of $x,\dot{x}$) is given by,
\be
\label{eq:penergy}
P(E) = \frac{\gamma \omega_0^2 - D}{2\omega_0^2\epsilon}
 \frac{1}{\left(1+\frac{D}{2\epsilon\omega_0^2}E\right)
^{\gamma \omega _0^2/D}}.
\ee
In the limit $D\rightarrow 0$ (purely additive driving), Eq.
(\ref{eq:penergy}) is replaced by,
\be
\label{eq:pex}
P(E) = \frac{\gamma}{\epsilon} \exp{(-\gamma E/\epsilon)}.
\ee
Equations (\ref{eq:penergy}) and (\ref{eq:pex})
clearly illustrate how the probability distribution of amplitude fluctuations
is qualitatively modified by the presence of multiplicative noise.
The dependence of $P(E)$ on $E$ is a power law in the multiplicative case
as opposed to the exponential decay for the additive case. As a consequence,
moments of $P(E)$ for the multiplicative case
diverge beyond an order that depends on $D$,
$\gamma$ and $\omega_0$, but not
on the additive noise intensity $\epsilon$. Hence,
in the white noise limit, there are always moments of sufficiently high order
that are unstable.
For narrow band noise, the stability boundaries of the various moments
are also expected to depend on the order of the moments considered.
However, the stability boundaries for all the various moments have to
converge to the stability diagram of the deterministic case,
as one narrows the width of the noise spectrum.

In summary, the stability of the solutions of Eq. (\ref{eq:osc})
is determined solely by the multiplicative
noise. Above threshold, the second order moments
grow exponentially in time and additive noise is irrelevant.
If all the interface modes are below threshold, which modes are
predominantly excited is determined by the dominant frequencies of the
additive noise. Higher order moments of the interface
displacement, however, may depend strongly on the parameters of the
multiplicative noise.

\section{Application to typical microgravity conditions}

Although a precise characterization of residual accelerations in a
microgravity environment is under way, there is enough information
already available for our purposes. The spectral density of g-jitter
determined during various space missions does have one or several
dominant frequencies, but it is also quite broad.

Contact with the results obtained for the
stochastic oscillator can be made in the case of a single fluid in an
infinite container
by replacing $\gamma = 4 \nu q^{2}$, $\omega_{0}^{2} \approx
\frac{\Gamma}{\rho} q^{3}$ and $\xi (t) = q g_{z}(t)$, where $q$ is the
wavenumber of the surface displacement away from planarity. Now $< \xi^{2} >
= q^{2} <g_{z}^{2} >$, with $<g_{z}^{2} >$ being proportional to the area
beneath the spectral density of g-jitter measured.
{}From Fig. 9 in ref. \cite{re:ne91} (which
gives the power spectral density of a representative time window aboard
Spacelab 3), we estimate $G = \sqrt{ \left< g_{z}^{2} \right> } \approx
8 \times 10^{-4} g_{E}$. This is a very conservative estimate;
considerably larger values can be obtained from this and other published
measurements (see, e.g., Figs. 1 and 2 in ref. \cite{re:al90}).
Note also that $\gamma$ and $\omega_0$ are no longer independent from
each other, implying that in the limit $\omega_0 \rightarrow 0$ the
system is actually approaching the underdamped limit,
$\gamma^2/4\omega_0^2 = 4 \nu^2 \rho \Gamma^{-1} q \ll 1$.
The stability boundary given by Eq. (\ref{eq:ansatz}) can be explicitly
written for this case as,
\be
\label{eq:neutralansatz}
\frac{G_{c}^{2}}{\nu \Omega^3}=4\frac{\omega_0^2}{\Omega^2}
\frac{1+2 (\Omega\tau)^2 \left(1+4
\left(\frac{\omega_0}{\Omega}\right)^2\right)+(\Omega\tau)^4
\left(1-4 \left(\frac{\omega_0}{\Omega}\right) ^2 \right)^2}
 {\Omega\tau \left[
1+(\Omega\tau)^2 \left(1+4
\left(\frac{\omega_0}{\Omega}\right) ^2\right)  \right]},
\ee
where $G_{c}$ is the critical value for instability.
Equation (\ref{eq:neutralansatz}) is plotted in Fig. \ref{fi:1} for several
representative values of $\Omega\tau$. The regions above
the curves correspond to regions of instability. The figure also shows
that there is a critical value of $\Omega\tau$ at which the curves change from
monotonic to non-monotonic behavior. The critical value is
$(\Omega\tau)_c \simeq 1.555$. The minimum at $\omega_0/\Omega=0$ is a
zero of the function and is independent of $\Omega\tau$.
The second minimum corresponds to the modes which resonate with the dominant
frequencies of the spectrum. For increasing $\Omega\tau$ this minimum
approaches $\omega_0 = \Omega/2$ from the left. The corresponding
value of $G_{c}^{2}/\nu\Omega^3$ decreases, as the spectrum becomes narrower
around $\Omega$ favoring the resonance.

For a given spectrum of g-jitter, there is always
a band of unstable modes at low frequencies (long wavelengths).
As we increase $\Omega\tau$
this band becomes narrower, and the critical wavelength for this low frequency
instability increases. For $\Omega\tau>(\Omega\tau)_c$ a second band of
unstable modes approximately centered at $\omega_0/\Omega=0.5$ may
appear depending on the level of g-jitter.
For the parameter ranges estimated for typical g-jitter, the low frequency
instability is likely to be unobservable because it may
correspond to wavelengths larger than the natural long wavelength
cut-off defined by the container size.
For instance, for fluid parameters appropriate to
a water-air interface,  $\nu = 0.01 cm^{2}/s, ~ \Gamma = 75.5 erg/cm^{2}$ and
$\rho =1g/cm^{3}$, with $\Omega \sim 12 Hz$ and $G \sim 10^{-3} g_E$
the critical wavelength ranges from $\sim 15 cm$ to $\sim 35 cm$ for
 $\Omega\tau$ in the range $1 - 25$.

Given that the low frequency band is likely not to be observable,
the absolute
threshold for instability will typically be given by the minimum of the curve
(\ref{eq:neutralansatz}). In Fig. \ref{fi:2} we have plotted the threshold
given by (\ref{eq:neutralansatz}) at $\omega=\Omega/2$, as a function
of $\Omega$ for several values of $\tau$ (the minimum of Eq.
(\ref{eq:neutralansatz}) is not exactly at $\omega=\Omega/2$, but the
value of $G^2/\nu\Omega^3$ at the exact minimum and at $\omega=\Omega/2$
do not differ significantly). The dashed portion of the curve corresponds
to the region where there is no resonance minimum ($\Omega \tau < (\Omega
\tau)_c$).
For comparison purposes, we have also plotted the deterministic case
given by $\xi(t) = \xi_{0} \cos \Omega t$
(the threshold for instability because of subharmonic resonance in this case
is given in Eq. (\ref{eq:detthres})). In order to compare
this case with the stochastic case, we have
$\left< \xi^{2} \right> = \xi_{0}^{2}/2$, i.e.,
$q^{2} \left< g_{z}^{2} \right> =
\xi_{0}^{2}/2$. The threshold for instability of the resonant mode
with $\omega_0=\Omega/2$ in this case is given by,
\begin{equation}
\label{eq:neutrald}
\left( G_c \right)_{det} = \nu \sqrt{2} \left(\frac{2 \rho \Omega^5}
    {\Gamma} \right)^{\frac{1}{3}}
\end{equation}
For the range of frequencies $5-20 Hz$ we see that a g-jitter level of
$G \simeq 10^{-3} g_E$ lies below all curves and therefore it does not
lead to instability.

Finally, Fig. \ref{fi:1} also shows that the interface becomes effectively
more stable as $\tau$
is decreased, at constant $G$ (i.e., at constant area of the power spectrum).
The forcing is less efficient in
exciting the resonance as it spreads into a wider band of frequencies,
as opposed to being concentrated at the resonant frequency.

\nonum
\section{Acknowledgments}

We thank Vlad Shapiro for many interesting discussions.
This work is supported by the Microgravity Science and Applications
Division of the NASA under contract No. NAG3-1284.
This work is also supported in part by the Supercomputer
Computations Research Institute, which is partially funded by the U.S.
Department of Energy, contract No. DE-FC05-85ER25000.

\newpage
\nonum
\section{Appendix}

  Here we discuss in some detail the estimation of the different contributions
to the total dissipation rate.

Viscous dissipation due to potential flow can be written as \cite{re:la59},
\be
\label{eq:dEdt}
{\cal D}_{bulk} = -2{\cal F}_{bulk} =
-2\mu_{1}\int_{V_{1}}\left(\frac{\partial^{2}\phi_{1}}
  {\partial x_{i}\partial x_{j}}\right)^{2}dV
-2\mu_{2}\int_{V_{2}}\left(\frac{\partial^{2}\phi_{2}}
  {\partial x_{i}\partial x_{j}}\right)^{2}dV,
\ee
where the integrals extend over the volume occupied by fluids 1 and 2
respectively.  When the nonlinear contributions in the interfacial
displacement are neglected, we find
\be
{\cal F}_{bulk} = (\mu_{1}+\mu_{2})
 \left(2S\sum_{n}\frac{k_{n}^{2}}{k_{n}'}\dot{h}_{n}^{2}(t)
 + \sum_{mn}\frac{\dot{h}_{m}\dot{h}_{n}}{k_{m}'k_{n}'}
 P_{mn} \int_{-d}^{0}dz Z_{1m}(z)Z_{1n}(z)\right)
\ee
where $P_{mn} = \int\!\!\int dxdy \nabla^{2}(\nabla\psi_{m} \cdot
\nabla\psi_{n})$.
The matrix $P_{mn}$ depends on the geometry of the
container and, in general, is neither zero nor diagonal.
Therefore the equation for $h_{n}(t)$ is coupled in general to all the
other modes $h_{m}(t)$. In the case of a rectangular
container of sides $L_{x}$ and $L_{y}$,
the eigenfunctions $\psi_{n}$ are
\be
\label{eq:eigen-rec}
\psi_{n}(x,y) \equiv \psi_{i,j}(x,y) =
\sqrt{(2-\delta_{i0})(2-\delta_{j0})}
\mbox{cos}\frac{i\pi x}{L_{x}}\mbox{cos}\frac{j\pi y}{L_{y}},
\ee
where $i,j=0,1,2,\cdots$, except $i=j=0$, and we have $P_{mn} = 0$ for
all $m,n$. For arbitrary geometry, $P_{mn}$ is expected to vanish
in the limit of large system size.

Dissipation in the viscous boundary layer near the container's wall
can be estimated by assuming a constant gradient of velocity in the
boundary layer for each eigenmode.  The thickness of the
boundary layer for mode $n$ is $\delta_{n} = \sqrt{2\nu/\omega_{n}}$,
with $\omega_{n}$ the characteristic angular frequency of $n-th$ mode. Then,
\be
{\cal D}_{wall} \sim
-(\rho_{1}\sqrt{\nu_{1}/2}+\rho_{2}\sqrt{\nu_{2}/2})
& & \sum_{mn}\frac{\sqrt{\omega_{m}}+\sqrt{\omega_{n}}}{2k_{m}'k_{n}'}f_{mn}
\dot{h}_{m} \dot{h}_{n},
\ee
where,
$$
f_{mn}=\int_{S_{1w}}\!\!dS \nabla(\psi_{m}Z_{1m}(z)) \cdot
 \nabla(\psi_{n}Z_{1n}(z))
=\int_{S_{2w}}\!\!dS \nabla(\psi_{m}Z_{2m}(z)) \cdot
 \nabla(\psi_{n}Z_{2n}(z)).
$$
$S_{1w}$ and $S_{2w}$ are the surfaces of the container wall
in contact with fluids 1 and 2 respectively.
The matrix $f_{mn}$ is not diagonal in general, and can be estimated to
be of order $\sim 2Lk_{m}k_{n}/(k_{m}+k_{n})$ with $L$ the lateral size
of the container.  With this estimation the ratio of bulk dissipation to
the dissipation near the container's wall is of the order of
$\sim Lk_{n}^{2}\sqrt{2\nu/\omega_{n}}$.

We now estimate dissipation in the boundary layer at the fluid interface
between two fluids.
Again, we assume a constant gradient of velocity inside the boundary layer for
each eigenmode, and equal to,
$|\phi_{1n}(t)\nabla(\psi_{n}Z_{1n})- \phi_{2n}(t)\nabla(\psi_{n}Z_{2n})|/
\delta_{n}$, with $\delta_{n}=\sqrt{2\nu/\omega_{n}}$.
Here $\nu$ and $\mu$ are the average kinematic and dynamic viscosities of
the two fluids.  Then we find,
\be
\label{eq:Dint}
{\cal D}_{int} \sim -4S\mu \sum_{n}\sqrt{\frac{\omega_{n}}{2\nu}}
\frac{k_{n}^{2}}{k_{n}^{'2}} \dot{h}_{n}^{2}.
\ee
This contribution dominates over bulk dissipation, even in the underdamped
limit and for a large container. The frequency dependence of
the resulting damping coefficient implies that a closed equation for
$h_{n}(t)$ cannot be obtained in the general case of a time dependent
gravitational field. If, on the other hand, the gravitational field is
constant, Eq. (\ref{eq:hn}) is approximately valid and contains a
damping coefficient that depends on the frequency of the mode
considered.

Finally, in the case of the free surface of a fluid ($\rho_2=0, \mu_2=0$),
the dissipation in the interface boundary layer can be neglected in front
of bulk dissipation, as discussed in Ref. \cite{re:zh92}.

\newpage

\figure{\label{fi:1}}  Stability boundaries for the second moments of
free surface away from planarity. The figure shows the dimensionless
mean squared fluctuations in gravitational acceleration versus the
dimensionless frequency of the surface modes. The intensity of the
fluctuations in the gravitational field is given by $G^{2} = <
g_{z}(t)^{2} >$, $\nu$ is the kinematic viscosity of the fluid, $\Omega$
the characteristic frequency of the narrow band noise, and $\omega_{0}$
the frequency of the surface mode. Different curves show the stability
boundaries for various values of the dimensionless correlation time
$\Omega \tau$. We note that for $\Omega \tau \ge 1.555$, the neutral
stability curve has one minimum at a finite value of
$\omega_{0}/\Omega $. Below that value, the only minimum is at
$\omega_{0} / \Omega = 0$.

\figure{\label{fi:2}} Estimate of tolerable levels of g-jitter
for instability of a planar water-air surface at room temperature. We
show the normalized root mean squared g-jitter for instability as a
function of the characteristic frequency of the driving noise ($g_{E}$
is
the intensity of the gravitational field on the Earth's surface). Three
different correlation times are shown, as indicated in the figure. The
solid lines refer to those regions of $\Omega \tau$ for which the
stability curve has a minimum at finite frequencies (see Fig.
(\ref{fi:1})). The
dotted lines represent the stability boundary at $\omega_{0} = \Omega
/2$, even though the stability curve does not have a minimum at this
point. The dashed line is the stability curve for the damped Mathieu equation
for the same driving frequency.

\end{document}